\documentclass[a4paper,10pt,oneside,twocolumn]{article}
\usepackage{amsmath}
\usepackage{amssymb}
\usepackage{units}
\usepackage[cp1250]{inputenc}
\usepackage[dvips]{graphicx}

\newcommand{\bfig}[1][t!]{\begin{figure}[#1] \begin{center}}
\newcommand{\efig}{\end{center} \rule{4cm}{0.4pt} \end{figure}}
\newcommand{\btab}[1][ht!]{\begin{table}[#1] \begin{center}}
\newcommand{\etab}{\end{center} \rule{4cm}{0.4pt} \end{table}}

\begin{document}

\title{Proton-Proton Physics with ALICE\footnote{The author thanks for support from the Institut f\"ur Kernphysik, Westf\"alische Wilhelms-Universit\"at M\"unster, Germany and the Istituto Nazionale di Fisica Nucleare, Sezione di Catania, Italy.}}

\author{Contribution for the 1st International Workshop on\\
Soft Physics
in ultrarelativistic Heavy Ion Collisions\\ \\Jan Fiete Grosse-Oetringhaus\\for the ALICE collaboration\\CERN, Switzerland}
\date{}

\maketitle

\section{Motivation}

ALICE -- A Large Ion Collider Experiment \cite{ppr1, ppr2} is a dedicated heavy-ion experiment at the LHC. Due to its detectors' characteristics it can study pp collisions as well as heavy-ion collisions. Special strength of ALICE is the low $p_{T}$ cut off ($\sim 0.1$~GeV/$c$) due to the low magnetic field, the small material budget ($X/X_0 \approx 10\%$)\footnote{For a track passing ITS and TPC.} and its excellent PID capabilities.
ALICE's minimum bias trigger, composed of signals from the V0 and the ITS pixels, is $\approx 88\%$ efficient for inelastic pp collisions \cite{trigger_claus}.

The motivation for analyzing pp events is threefold: A (more technical) motivation is that pp collisions are optimal for commissioning of the detector due to their low multiplicity. Most of the calibration and alignment tasks can be performed more efficiently with low multiplicity events. Furthermore the study of pA and AA collisions requires a pp reference at the same energy. Obtaining the latter in the same detector minimizes systematic effects.
The main motivation, however, is the study of pp collisions at a new energy. The LHC will allow for the first time to analyze physics above the "knee" in the cosmic-ray energy distribution\footnote{The flux of cosmic rays as a function of the cosmic ray energy shows an exponential behavior. The slope changes between $10^{15}$ -- $10^{16}$ eV, which, due to the shape in a double-logarithmic scale, is called knee.}. This interesting insight in a new energy regime is especially important for the tuning of Monte Carlo predictions. These are needed in further measurements to be able to distinguish rare processes from the underlying event.

\bfig
	\includegraphics[width=\columnwidth]{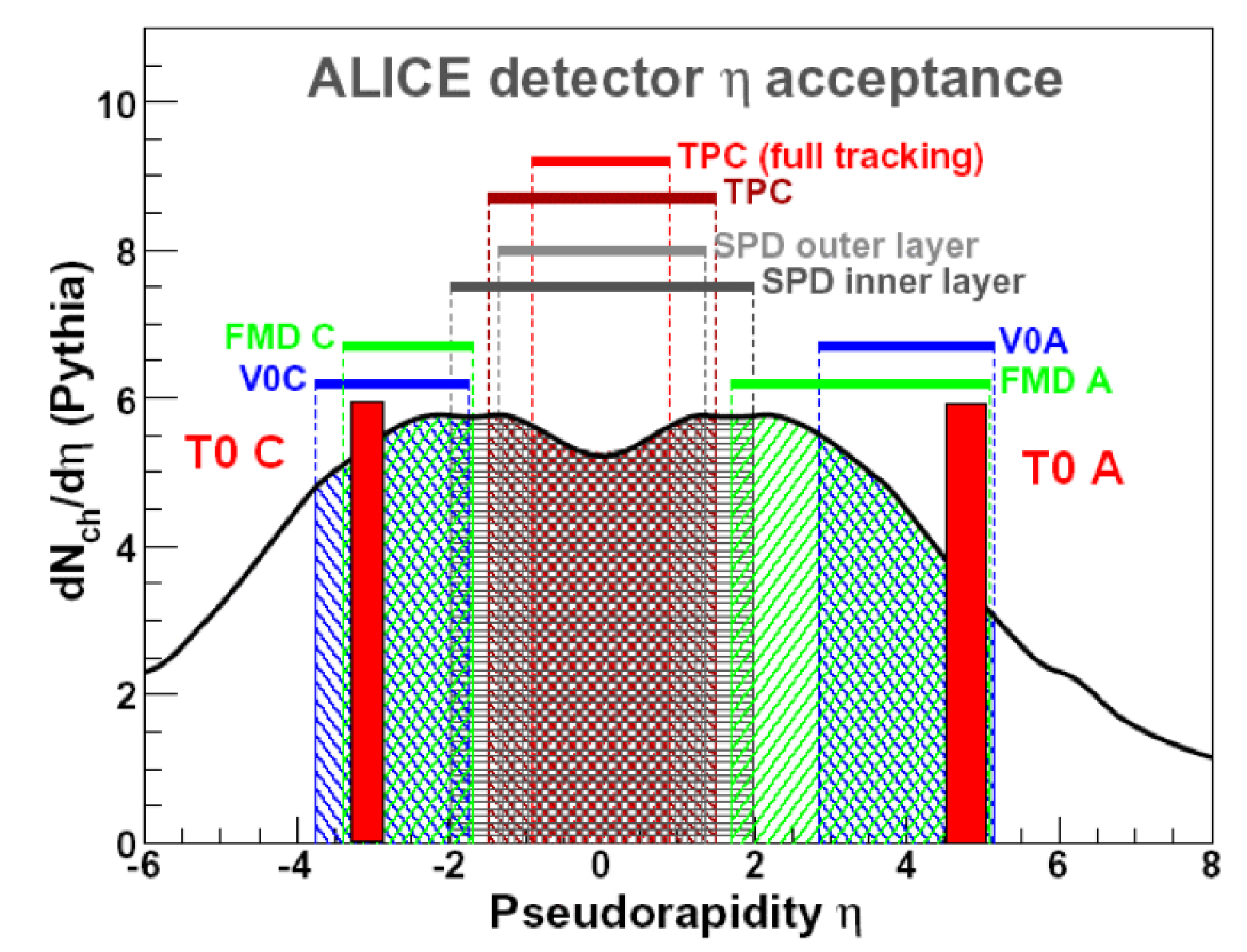}
  \caption{Pseudorapidity acceptance for charged particles (without muons) superimposed with the Pythia prediction. Shown is the reach of several detector systems. Please note that not all systems provide a momentum measurement.}
  \label{acceptance}
\efig

\section{The ALICE detector}

ALICE is designed to be able to reconstruct events with a charged track multiplicity at mid-rapidity ($dN_{ch}/d\eta\ |\ \eta=0$) up to 8000. However, w.r.t. the measurements at RHIC a lower value of 1800 -- 2000 for central collisions is predicted \cite{mult}.

The detector system at central rapidity is able to identify hadrons, electrons and
photons from very low transverse momenta around 100 MeV/$c$, to large momenta of 100 GeV/$c$.
It consists of the Inner Tracking System (ITS) featuring six layers of
high-resolution silicon detectors, the Time Projection Chamber (TPC) as the main
tracking system of the experiment, the Transition Radiation Detector (TRD) which
provides electron identification, and the Time Of Flight (TOF) detector for particle
identification. These detectors have full azimuthal and central rapidity ($|\eta| < 0.9$)
coverage. The design also includes two small-area detectors: an array of ring-imaging
Cherenkov detectors for identification of high-momentum particles (High-Momentum
Particle Identification Detector -- HMPID) and an electromagnetic calorimeter made
of high density crystals (Photon Spectrometer -- PHOS). The central barrel is covered
by a magnetic field of a maximum 0.5 T.

At large rapidities the detector system consists of a muon spectrometer shielded by an
absorber, and multiplicity detectors -- the Forward Multiplicity Detector (FMD) and
the Photon Multiplicity Detector (PMD). Trigger signals are provided by scintillators
and quartz counters (T0 and V0). The Zero-Degree Calorimeter (ZDC) provides information that can be used
to measure the impact parameter. It consists of two sets of neutron and hadron calorimeters
which are located about 100~m away from the interaction point.

\section{LHC Commissioning}

The LHC commissioning scenario as it is known by today \cite{lhc_comm_slides} foresees the closure of the collider at the end of August 2007. The machine will then be operated for 2 -- 3 months providing beam-gas events, first collisions at $\sqrt{s}$ = 900 GeV and possibly at $\sqrt{s}$ = 2.2 -- 2.4 TeV. Subsequently a shutdown from January -- March 2008 is planned.

The general guideline is to achieve the highest possible energy as soon as possible at low luminosity ($< 6 \cdot 10^{28}$~cm$^{-2}$~s$^{-1}$), which is a unique opportunity for ALICE to measure pp collisions. When the LHC reaches nominal luminosity of $10^{34}$~cm$^{-2}$~s$^{-1}$ new strategies will be needed to be able to still record meaningful pp data due to the occurring pile-up (e.g. beam displacement, defocussing at point 2).

During commissioning ALICE will be able to record about 3.6 million events per day\footnote{Assuming 10 hours per day at a DAQ rate of 100 Hz.}.

\section{Physics topics}

\bfig
	\includegraphics[width=\columnwidth]{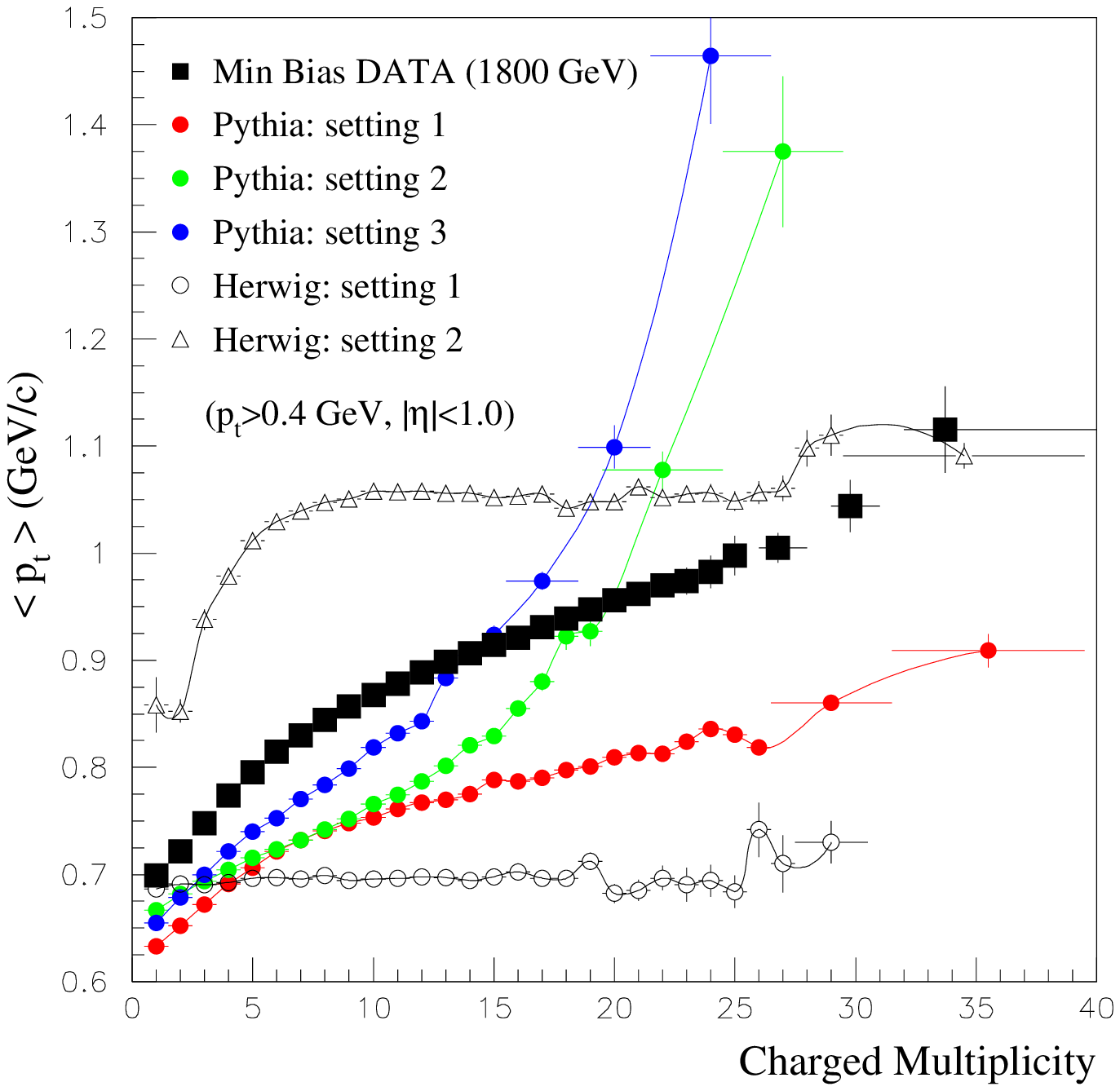}
	\caption{$\left<p_{T}\right>$ vs. charged multiplicity measured by CDF \cite{cdf_ptvsmult}. Black squares are data points, the other curves are predictions by Pythia and Herwig with different tunes. Note that none of them reproduces the data.}
	\label{ptvsmult}
\efig

Soft physics topics range from pseudorapidity, multiplicity and $p_T$ distributions of charged particles to identified particle spectra, strangeness production, resonance production, correlations and baryon transport. Some of the topics which can be addressed during the first days of data taking will be outlined in this section (see also \cite{alice_first_physics}).

\paragraph{$dN_{ch}/d\eta$ distribution}
The pseudorapidity rapidity density at mid-rapidity ($dN_{ch}/d\eta | \eta = 0$) was expected to follow logarithmic "Feynman" scaling ($\sim ln(s)$) \cite{feynman_scaling}, which was broken at SPS energies \cite{Arnison:1982rm}. Current phenomenological expectations assume quadratic logarithmic scaling ($\sim ln^2(s)$) (e.g. \cite{Abe:1989td}). ALICE can measure the pseudorapidity distribution in the central region using two detectors: The ITS pixels and the TPC. In the forward region the measurement is performed with the FMD. The number of needed events is quite small, already 1,000 events lead to a statistical error of about 3\%.

\paragraph{The multiplicity distribution} is the probability $P_{N}$ that an event has multiplicity N. Measurements are well described by KNO scaling until SPS energies \cite{kno_sps}. The breaking of KNO scaling is contributed to correlated particle emission. Current predictions use negative binomial distributions \cite{binom} or a two component approach \cite{twocomponent1}. ALICE's reach in multiplicity with $10^7$ events is about 125 ($|\eta| < 0.9$) \cite{ppr2}.

\paragraph{The $p_{T}$ spectrum} can be described in the soft region by a phenomenological approach, in the hard region by N(LO) pQCD\footnote{Although pQCD compared to a phenomenological approach seems more exact, it still contains uncertainties in the choice of the parton distribution and fragmentation functions.}. ALICE's reach in $p_T$ with $10^7$ events is about 30 GeV/$c$ ($|\eta| < 0.9$) \cite{ppr2}.\footnote{The $p_T$ reach is strongly dependent on the binning. In the referenced study 100 tracks above the value given (30 GeV/$c$) are required.}

\paragraph{$\left<p_T\right>$ vs. multiplicity} describes the balance between particle production and transverse energy. It is found that $\left<p_T\right>$ increases with multiplicity, which is attributed to jets. However CDF showed that this is also seen in events without jets \cite{cdf_ptvsmult}. This behavior is not yet satisfactory explained by models or MC generators (see Figure \ref{ptvsmult}).

\section{First Physics Preparation}

The analysis tools for first physics topics are already being prepared and their various systematic effects estimated. For this purpose a physics data challenge is ongoing that is producing $10^8$ simulated pp events. Three different event samples are created that resemble the detector in different conditions w.r.t. the alignment status: 1) the perfectly aligned detector, 2) the misaligned detector as expected before applying alignment procedures (this will be the situation at the beginning of data taking) and 3) the misaligned detector after applying alignment procedures.

On the basis of this data the analysis code is developed. As an example the $dN_{ch}/d\eta$ measurement is outlined in the next section.

\subsection{$dN_{ch}/d\eta$ measurement}

\bfig
	\includegraphics[width=\columnwidth]{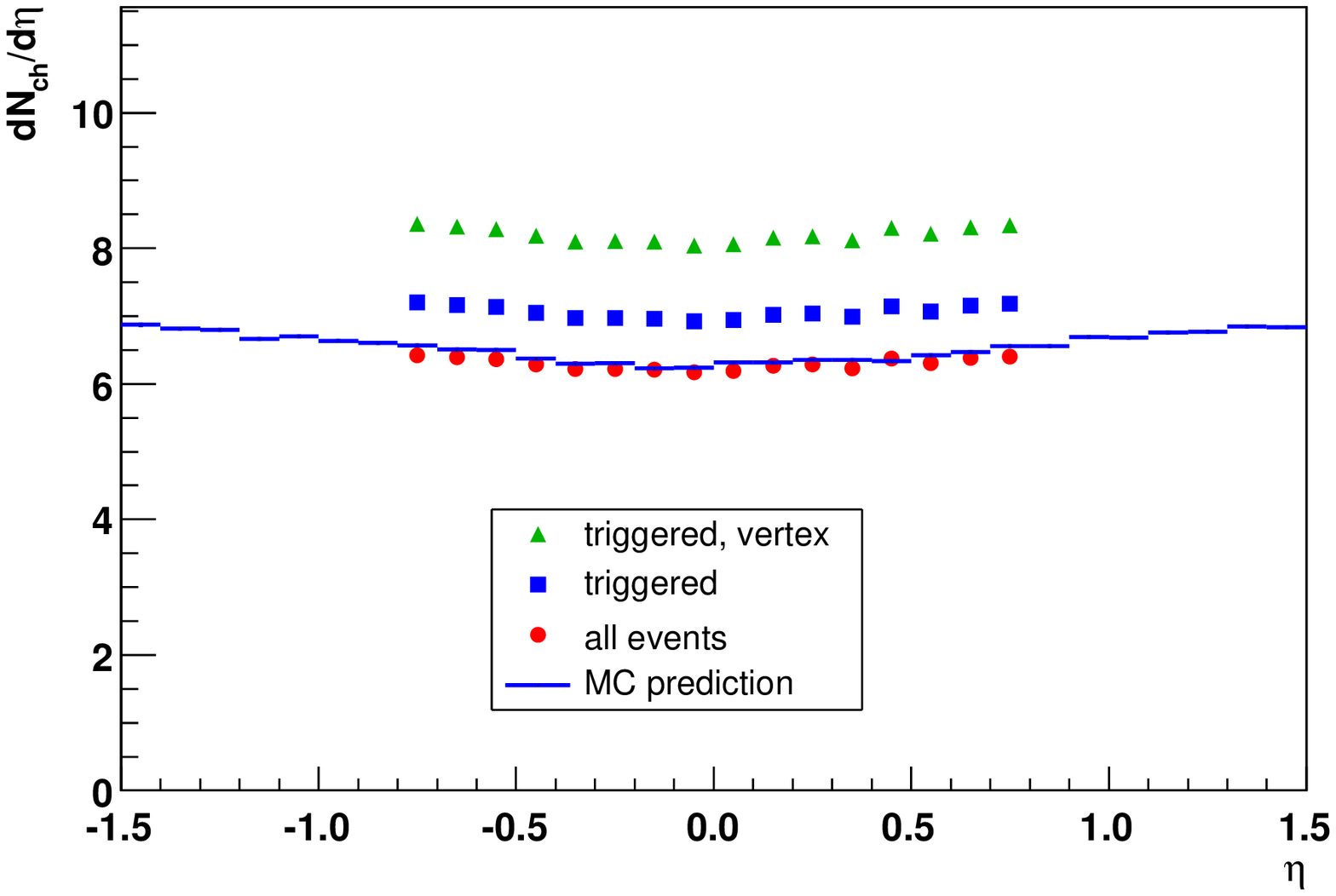}
	\caption{The result of the analysis
    together with the input (MC) distribution. Three different
    $dN_{ch}/d\eta$ distributions are shown: 1) for all
    events that are triggered and have a reconstructed vertex (triangles), 2) all triggered events (squares) and
    3) all events (circles).}
	\label{dndeta}
\efig

The procedure is straightforward: First certain quality criteria is applied on event and track level. E.g. only events with reconstructed vertex are considered and only tracks with a reasonable $\chi^2$ and originating from the primary vertex. The tracks are counted and three corrections are applied: The first corrects for effects of geometrical acceptance, reconstruction efficiency, decay and feed-down. It is applied on track-level.
The second correction takes into account the bias introduced by the
  vertex reconstruction on the triggered event sample.
The third correction corrects for the trigger bias. The latter two corrections are applied on event level.
  Figure \ref{dndeta} shows the result of the analysis of simulated data.

\section{Summary}

ALICE has unique capabilities for pp physics that will lead to many interesting measurements at a new energy. The first data-taking period promises an event sample of the order of $\ge 10^7$ events that should allow to exceed the reach of previous measurements by far. The preparation of the tools has already started in order to be ready for the first collision.

The author would like to thank Claus J\o rgensen for many fruitful discussions and good teamwork.

\end{document}